\documentclass{mpe_report}
\usepackage{psfig,graphics,epsfig,lscape,txfonts,color}
\usepackage{natbib}
\bibpunct{(}{)}{;}{a}{}{,}

\def\cm-2{cm$^{-2}$}

\def\ein{{\it Einstein}}
\def\chandra{{\it Chandra}}

\def\xmm{{XMM-Newton}}
\def\m31{\object{M~31}}

\newcommand{\hcm}[1]{$\times 10^{#1}$ cm$^{-2}$}

\newcommand{\nh}{\hbox{$N_{\rm H}$}}

\newcommand{\mr}{\mathrm}

\begin{document}
\title{A deep XMM-Newton survey of M~31\thanks{\footnotesize Based on observations obtained with XMM-Newton, an ESA science mission with instruments and contributions directly funded by ESA Member States and NASA.}}
\author{H.~Stiele\inst{1}, W.~Pietsch\inst{1},  F.~Haberl\inst{1}, R.~Barnard\inst{2}, V.~Burwitz\inst{1}, M.~Freyberg\inst{1}, J.~Greiner\inst{1}, D.~Hatzidimitriou\inst{3}, M.~Hernanz\inst{4}, U.~Kolb\inst{2}, A.~Kong\inst{6}, P.~Plucinsky\inst{7}, P.~Reig\inst{8}, M.~Sasaki\inst{7}, G.~Sala\inst{1}, L.~Shaw Greening\inst{2}, L.~Stella\inst{5} \and B.~Williams\inst{9}}  
\institute{Max--Planck--Institut f\"ur extraterrestrische Physik, Giessenbachstra{\ss}e, 85748 Garching, Germany
\and Department of Physics and Astronomy, The Open University, Walton Hall, Milton Keynes, MK7 6AA, UK
\and University of Crete, Department of Physics, PO Box 2208, 71003 Heraklion, Greece
\and Institut de Ci\`encies de l'Espai (CSIC-IEEC), E-08193 Bellaterra (Barcelona), Spain
\and INAF-Osservatorio Astronomico di Roma, Via Franscati 33, I-00040 Monteporzio Catone, Roma, Italy
\and Institute of Astronomy, National Tsing Hua University, Hsinchu, Taiwan
\and Harvard-Smithsonian Center for Astrophysics, 60 Garden Street, Cambridge, MA 02138
\and IESL, Foundation for Research and Technology, 71110 Heraklion, Greece
\and Department of Astronomy and Astrophysics, Pennsylvania State University, University Park, PA}
\offprints{\footnotesize H.~Stiele, email: hstiele@mpe.mpg.de}
\titlerunning{A deep XMM-Newton survey of M~31}
\authorrunning{H.~Stiele et al.}
\maketitle

\begin{abstract}
The deep homogeneous survey of the large Local-Group spiral galaxy M~31 is a milestone project for X-ray astronomy, as it allows a detailed X-ray inventory of an archetypal low-star-formation-rate galaxy like our own. We present first results of the survey, which covers the entire $\mathrm{D_{25}}$ ellipse. Information from different X-ray energy bands are combined in an X-ray colour image of M~31. In the first 15 observations we found about 1000 sources, the full survey will yield about 2000 X-ray sources. \\
Sources will be classified using hardness ratios, extent, high quality spectra and time variability. In addition the sources will be correlated with catalogues in optical, infra-red and radio wavelengths.\\
Our goal is to study M~31 X-ray binaries and globular cluster sources, supersoft sources, supernova remnants and the hot interstellar medium and separate them from foreground stars and background objects.   
\end{abstract}

\section{Introduction}
The goal of this survey is to improve our understanding of the different source classes in \m31\ and possible connections to the general structure of the galaxy. Additionally the interstellar medium (ISM) will be studied. The results achieved by this survey will help us to better understand the source classes of the Milky Way and of more distant galaxies. Due to its relatively short distance \citep[780 kpc, ][]{1998AJ....115.1916H,1998ApJ...503L.131S} and its moderate foreground absorption \citep[\nh = 7\hcm{20}, ][]{1992ApJS...79...77S}, the bright, massive spiral (SA(s)b) galaxy \m31\ is an ideal target for such a study. \\
Early observations in the X-ray wavelength regime were performed by the \ein\ observatory \citep[][]{1979ApJ...234L..45V,1990ApJ...356..119C,1991ApJ...382...82T} and ROSAT \citep[][]{1993ApJ...410..615P,1997A&A...317..328S,2001A&A...373...63S}. In the last few years many observations with \chandra, \xmm, Swift and other X-ray satellites covered mainly the central region of \m31. Individual references will be given in Section~\ref{Ssclass}, where the different source classes are discussed. \chandra\ results are discussed in a different paper of this meeting \citep{2007Chan..M31....1W}.     

\section{Observations and data}
\begin{figure}
\centerline{\psfig{file=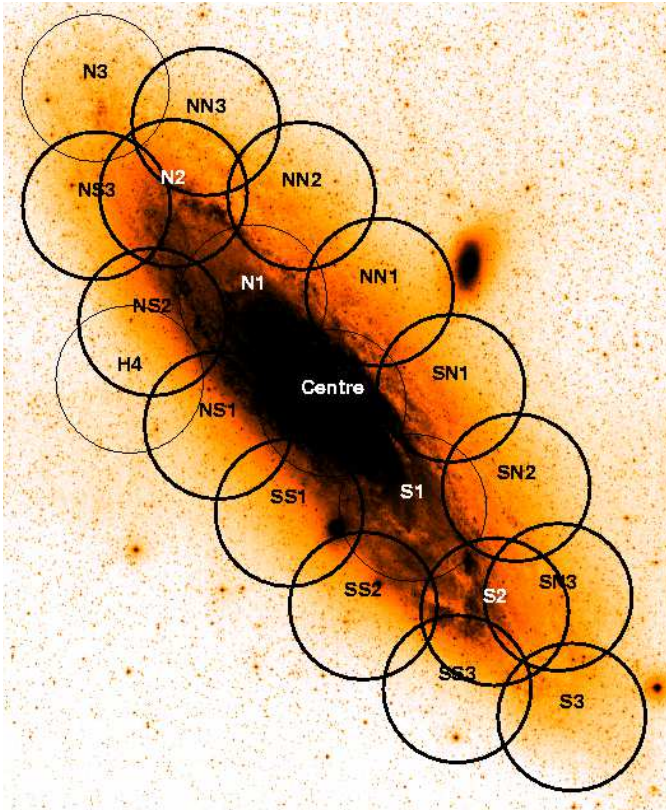,width=8.8cm,clip=} }
\caption{This image shows the pointings of the 15 observations of the large survey overlaid on a deep optical image of \m31. The region covered by a single observation is marked with a bold circle with a radius of 14 arcmin.  
\label{fields}}
\end{figure}

The survey consists of 15 individual EPIC observations. They are placed in such a way that the whole $\mathrm{D_{25}}$ ellipse is covered (see Fig.~\ref{fields}). The observations were taken in \xmm\ AO~5, between June 2006 and January 2007.\\
For each observation a high energy (7 to 15~keV) background light curve was produced. We used these light curves to select time intervals with low background, the so-called good time intervals (GTIs). As many of the observations were affected by strong background flares (mostly at the beginning or end of the observation), the net exposure useful for our analysis is strongly reduced. The sum of GTIs after high energy background screening ranges from 6 to about 55 ks. The observations taken during the summer visibility window of \m31\ were affected more strongly by background radiation than those taken during the winter window. The most affected observations were reapproved in \xmm\ AO~6. A severe background screening is especially important for the detection of faint point sources, investigation of diffuse emission and for colour images. We also want to mention, that for some of the observations the soft energy ($< 2$~keV) background varied and an additional screening was necessary.\\
We used five energy bands: (0.2--0.5) keV, (0.5--1.0) keV, (1.0--2.0) keV, (2.0--4.5) keV, and (4.5--12) keV, to create images, background images and exposure maps for PN, MOS\,1 and MOS\,2 and masked them for acceptable detector area. For PN the background maps contain the contribution from the ``out of time (OOT)" events.\\
Source detection is done on all 15 images simultaneously, using the \xmm\ Science Analysis System (SAS) tasks {\tt eboxdetect} and {\tt emldetect}. Details can be found e.\,g.~in \citet{2007TV....M31....1S}. We detected about 1000 sources in the AO5 observations. Together with the sources found by \citet{2005A&A...434..483P} and the reapproved observations, the \xmm\ source catalogue will contain about 2000 individual sources in the field of \m31. 

\section{Colour image}
\label{Scoim}
\begin{figure}
\centerline{\psfig{file=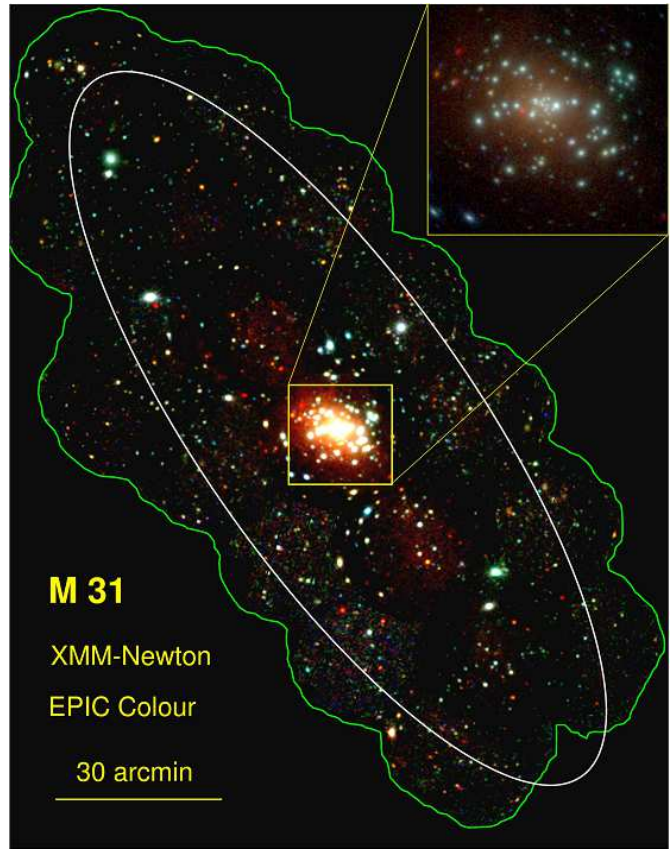,width=8.8cm,clip=} }
\caption{First combined EPIC PN, MOS\,1 and MOS\,2 RGB image of the new and archival data. For more details see Sec.~\ref{Scoim}.
\label{cimage}}
\end{figure}

Figure~\ref{cimage} shows a first combined exposure corrected EPIC PN, MOS\,1 and MOS\,2 RGB image of the new and archival data. The colours represent the X-ray energies as follows: red $0.2 - 1$ keV, green $1 - 2$ keV and blue $2 - 12$ keV. The optical extent of \m31\ is indicated by the $\mathrm{D_{25}}$ ellipse and the boundary of the observed field is given by the green contour. The image is smoothed with a Gaussian of 20" FWHM. Observation S3 was removed completely due to strong background. In some observations individual noisy MOS\,1 and MOS\,2 CCDs are omitted.\\
The colour of the sources reflects their class. Supersoft sources (SSSs) appear in red. Thermal supernova remnants (SNRs) and foreground stars (fg stars) are red to yellow. "Hard" sources (background objects (AGNs), X-ray binaries (XRBs) or Crab-like SNRs) are blue. 

\section{Source classes}
\label{Ssclass}
We want to classify and if possible identify the sources of our survey. In this way we will separate sources belonging to \m31\ from foreground stars and background objects. In the following we will discuss the different methods, which help us to differentiate the source populations.  

\subsection{Hardness ratios}
\begin{figure}
\centerline{\psfig{file=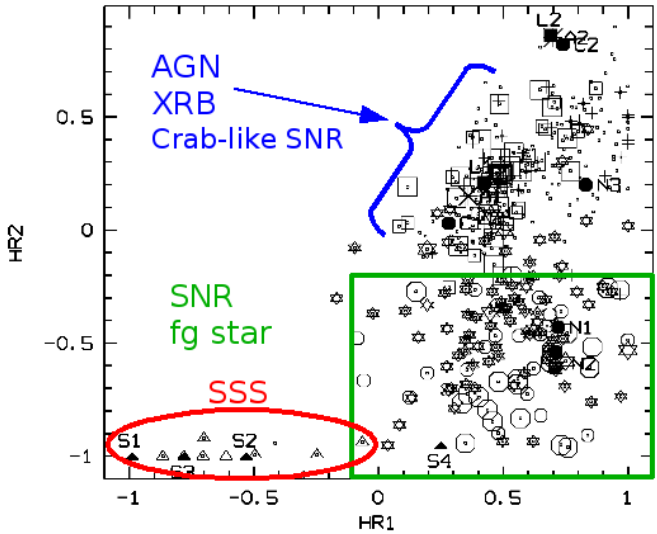,width=8.8cm,clip=} }
\caption{Hardness ratio plot HR2 versus HR1 for the sources of the catalogue of \citet{2005A&A...434..483P}.\@ Marked are the three regions, which separate clearly from each other. Source classification is indicated: Foreground stars and candidates are marked as big and small stars, AGN candidates as small crosses, SSS candidates as triangles, SNR and candidates as big and small hexagons, GlCs and XRBs and candidates as big and small squares.
\label{HRplot}}
\end{figure}

The most general method, which can be applied to all sources and which is based on their X-ray spectral properties, is the analysis of hardness ratios. The hardness ratios and errors are defined as  
\begin{equation}
HRi = \frac{B_{i+1} - B_{i}}{B_{i+1} + B_{i}}\; \mbox{and}\;\; EHRi = 2  \frac{\sqrt{(B_{i+1} EB_{i})^2 + (B_{i} EB_{i+1})^2}}{(B_{i+1} + B_{i})^2},
\end{equation}
for {\it i} = 1 to 4, where $B_{i}$ and $EB_{i}$ denotes count rates and corresponding errors in band {\it i} as defined above. A hardness ratio can be recognised as an X-ray colour.\\
Figure~\ref{HRplot} shows HR2 plotted versus HR1 for sources of the catalogue of \citet{2005A&A...434..483P}.\@ The sources group in three regions, which separate clearly. In the lower left corner SSSs can be found (HR1 $< 0$ and HR2 $\sim -1$). The major class of SSSs in \m31\ are optical novae. This was found by \citet{2005A&A...442..879P} correlating SSSs detected with ROSAT, \chandra\ and \xmm\ with optical nova catalogues from the literature and from their own observations. The project is continued and recent results are published in \citet{2007A&A...465..375P}. Two SSSs which are not associated with optical novae are known to show periodicities of 865 s and 217 s, respectively. They are suggested to be spinning white dwarfs in a post-nova state \citep{2001A&A...378..800O,2007arXiv0708.0874T}. \\
SNRs and fg stars are located in the lower right corner of that hardness ratio plot (HR1 $> -0.1$ and HR2 $< -0.2$). All "hard" sources (AGNs, XRBs and Crab-like SNRs) can be found in the upper right.

\subsection{Time variability}
To further distinguish these source classes, an investigation of their X-ray time variability is important. SNRs do not show time variability of their overall intensity on time scales of years. 
In this respect SNRs and fg stars can be separated which is not possible from HRs alone. A fg star classification can be supported by an optical counterpart or if the star also shows variability on short timescales. A reason for this variability can be flares (see Fig.~\ref{flstar} and below). \\ 
With the help of time variability one can also separate XRBs from AGNs and Crab-like SNRs. Only XRBs can show strong time variability. Among the AGNs we know only a handful in the whole sky which show variability comparable to that of XRBs. So it is very unlikely that a strongly ($F_{\mr{var}} > 10$) variable source in the field of \m31\ would be an AGN. A detailed study on the central region of \m31\ is presented by \citet{2007TV....M31....1S}.\\
In addition XRBs can be transient sources. There are many studies of transient sources in \m31\, like e.\,g.~\citet{2006ApJ...643..356W}, \citet[][]{2006ApJ...645..277T}, \citet{2005ApJ...632.1086W}, \citet[][]{2006ApJ...637..479W}. A transient which we examined within our survey is discussed in Section~\ref{Strans}. It should be mentioned that up to now among XRBs only low mass systems are known in \m31. 

\subsection{Cross correlations}
So far only X-ray properties of the sources have been used for source classification. Additional information can be achieved using cross correlations with catalogues in the radio, infra-red and  optical wavelengths. X-ray sources which correlate with optical globular clusters are most likely X-ray binaries. Bright XRBs in globular clusters were investigated by \citet{2002ApJ...570..618D}. Optical images can be helpful to distinguish between fg stars and SNRs. Correlations with radio counterparts can support AGN and SNR classifications. For the classification of AGNs also optical spectra can be helpful. In \chandra\ observations SNRs can be found as spatially resolved X-ray sources \citep{2003ApJ...590L..21K}.\@ So the extent of an X-ray source can be an additional criterion for a SNR classification.

\subsection{Additional methods for bright sources}
\begin{figure}
\centerline{\psfig{file=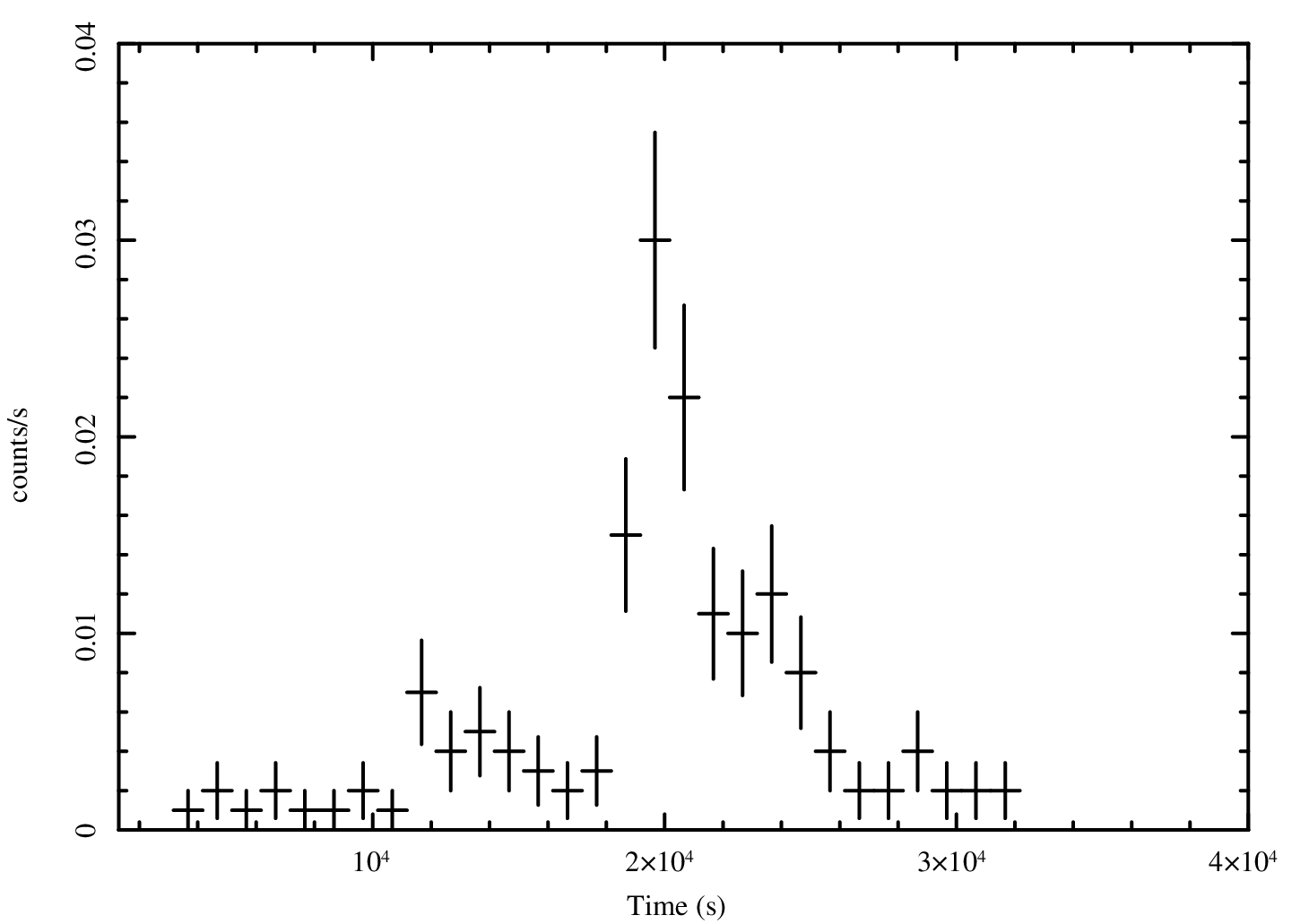,width=6.2cm,clip=,angle=-90} }
\caption{Combined EPIC PN and MOS 1000\,s light curve of a bright source showing a flare. Corresponding counterparts are found in the Local Group Survey \citep{2006AJ....131.2478M}, 2MASS and USNO-B1 catalogues. Thus it can be classified as fg star. This source was also detected with \chandra\ \citep[][ n1-15]{2004ApJ...610..247D} and ROSAT \citep[][ \# 327]{1997A&A...317..328S}.
\label{flstar}}
\end{figure}

We examined 1000\,s light curves of all bright objects in our survey for flares. An example is given in Figure~\ref{flstar}.\@ If in addition an optical counterpart with low ratio of the X-ray to the optical (V band) flux ($\log(f_{\mr{x}}/f_{\mr{opt}}) < -1.0$) is found, the source must be a fg star. To clearly see the rise and decline of the flare it is necessary to have a long and continuous observation.\\
The classification of brighter sources can be improved using X-ray spectra and short term time variability.

\section{A new X-ray transient}
\label{Strans}
\begin{figure}
\centerline{\psfig{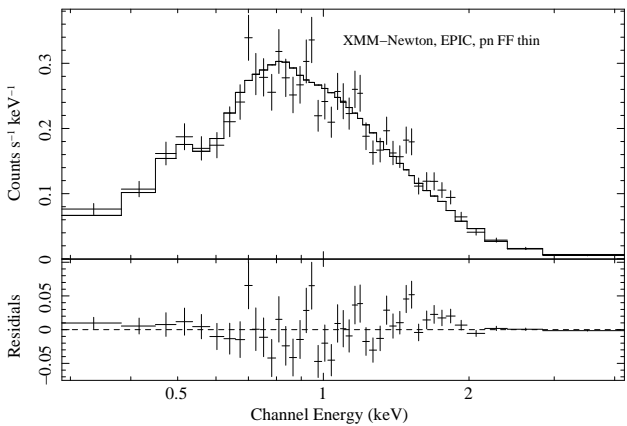} }
\caption{$0.2 - 7.0$~keV EPIC PN spectrum of the transient CXOM31 J004059.2+411551, fitted by an absorbed disk blackbody model. Best fit values given in Sec.~\ref{Strans}.
\label{Tspec}}
\end{figure}

The transient (CXOM31 J004059.2+411551) was detected with \chandra\ on 2007 July 5 \citep{2007ATel.1147....1G}. The position of the transient nearly coincides with the centre of the SN\,1 pointing of our survey. During a 22\,ks \xmm\ target of opportunity observation (TOO) on 2007 July 25, about 20 days after the \chandra\ detection, the source was still bright. We fitted the $0.2 - 7.0$~keV EPIC PN spectrum (see Fig~\ref{Tspec}) by an absorbed disk blackbody model with best fit parameters $N_{\mr{H}} = 2.1 \pm 0.2 \times 10^{21} \mr{atoms/cm^2}$ and $T_{\mr{in}}=0.51 \pm 0.02$~keV.\@ The unabsorbed luminosity ($0.5 - 10$~keV) is $L_{\mr{x}}\sim9.1\times10^{37}\mr{erg/s}$.\@ The spectral parameters and luminosity are in agreement with the \chandra\ values. A FFT period search did not reveal any significant periodicities in the 0.3 to 2000 s range. We did not find any evidence of optical counterparts for the transient, checking the \xmm\ optical monitor UVW1, UVM2 images, taken during the TOO observation, and the Local Group Survey \m31 images \citep{2006AJ....131.2478M}. The X-ray parameters and the lack of an optical counterpart are consistent with this source being a black hole X-ray transient \citep{2007ATel.1191....1S}.

\section{Summary}
The deep \xmm\ survey of \m31\ consists of fifteen pointing directions, which are placed in such a way that together with the observations in the \xmm\ archive the whole $\mathrm{D_{25}}$ ellipse of the galaxy is covered.  The observations were taken in AO5 between June 2006 and January 2007. Some of the observations affected by high background were reapproved in AO6. The full survey will yield about 2000 sources in the field of \m31. This survey will separate sources belonging to \m31\ (supersoft sources, quasisoft sources, supernova remnants and X-ray binaries) from background objects, especially AGNs, and foreground stars.\\
We produced an RGB colour image in which SSSs appear as red, fg stars and SNRs as red to yellow sources. All "hard" sources are blue. A more qualitative form of the selection criterion, which forms the basis of this colour selection, are hardness ratios. In addition X-ray time variability can be helpful to distinguish between different source classes. Further information is achieved by cross correlations with catalogues in other wavelengths. For bright sources X-ray light curves, X-ray spectra and short term time variability can be examined.  \\
The next steps will be an improvement of the source positions and the preparation of a source catalogue.

\begin{acknowledgements}
\scriptsize
\footnotesize
The XMM-Newton project is supported by the Bundesministerium f\"ur Wirtschaft und Technologie/Deutsches Zentrum f\"ur Luft- und Raumfahrt (BMWI/DLR, FKZ 50 OX 0001) and the Max-Planck Society. H.~Stiele acknowledges support by the Bundesministerium f\"ur Wirtschaft und Technologie/Deutsches Zentrum f\"ur Luft- und Raumfahrt (BMWI/DLR, FKZ 50 OR 0405).
\end{acknowledgements}

\bibliographystyle{aa}
\bibliography{./paper,/home/hstiele/data1/papers/my1990,/home/hstiele/data1/papers/my2000,/home/hstiele/data1/papers/my2001,/home/hstiele/data1/papers/catalog,/home/hstiele/data1/papers/my2007}

\begin{thebibliography}{26}
\small
\expandafter\ifx\csname natexlab\endcsname\relax\def\natexlab#1{#1}\fi

\bibitem[{{Collura} {et~al.}(1990){Collura}, {Reale}, \&
  {Peres}}]{1990ApJ...356..119C}
{Collura}, A., {Reale}, F., \& {Peres}, G. 1990, ApJ, 356, 119

\bibitem[{{Di Stefano} {et~al.}(2002){Di Stefano}, {Kong}, {Garcia}, {Barmby},
  {Greiner}, {Murray}, \& {Primini}}]{2002ApJ...570..618D}
{Di Stefano}, R., {Kong}, A.~K.~H., {Garcia}, M.~R., {et~al.} 2002, ApJ, 570,
  618

\bibitem[{{Di Stefano} {et~al.}(2004){Di Stefano}, {Kong}, {Greiner},
  {Primini}, {Garcia}, {Barmby}, {Massey}, {Hodge}, {Williams}, {Murray},
  {Curry}, \& {Russo}}]{2004ApJ...610..247D}
{Di Stefano}, R., {Kong}, A.~K.~H., {Greiner}, J., {et~al.} 2004, ApJ, 610,
  247

\bibitem[{{Galache} {et~al.}(2007){Galache}, {Garcia}, {Steeghs}, {Torres},
  {Murray}, {Primini}, \& {Williams}}]{2007ATel.1147....1G}
{Galache}, J.~L., {Garcia}, M.~R., {Steeghs}, D., {et~al.} 2007, The
  Astronomer's Telegram, 1147, 1

\bibitem[{{Holland}(1998)}]{1998AJ....115.1916H}
{Holland}, S. 1998, AJ, 115, 1916

\bibitem[{{Kong} {et~al.}(2003){Kong}, {Sjouwerman}, {Williams}, {Garcia}, \&
  {Dickel}}]{2003ApJ...590L..21K}
{Kong}, A.~K.~H., {Sjouwerman}, L.~O., {Williams}, B.~F., {Garcia}, M.~R., \&
  {Dickel}, J.~R. 2003, ApJ, 590, L21

\bibitem[{{Massey} {et~al.}(2006){Massey}, {Olsen}, {Hodge}, {Strong},
  {Jacoby}, {Schlingman}, \& {Smith}}]{2006AJ....131.2478M}
{Massey}, P., {Olsen}, K.~A.~G., {Hodge}, P.~W., {et~al.} 2006, AJ, 131, 2478

\bibitem[{{Osborne} {et~al.}(2001){Osborne}, {Borozdin}, {Trudolyubov},
  {Priedhorsky}, {Soria}, {Shirey}, {Hayter}, {La Palombara}, {Mason},
  {Molendi}, {Paerels}, {Pietsch}, {Read}, {Tiengo}, {Watson}, \&
  {West}}]{2001A&A...378..800O}
{Osborne}, J.~P., {Borozdin}, K.~N., {Trudolyubov}, S.~P., {et~al.} 2001, A\&A,
  378, 800

\bibitem[{{Pietsch} {et~al.}(2005{\natexlab{a}}){Pietsch}, {Fliri}, {Freyberg},
  {Greiner}, {Haberl}, {Riffeser}, \& {Sala}}]{2005A&A...442..879P}
{Pietsch}, W., {Fliri}, J., {Freyberg}, M.~J., {et~al.} 2005{\natexlab{a}},
  A\&A, 442, 879

\bibitem[{{Pietsch} {et~al.}(2005{\natexlab{b}}){Pietsch}, {Freyberg}, \&
  {Haberl}}]{2005A&A...434..483P}
{Pietsch}, W., {Freyberg}, M., \& {Haberl}, F. 2005{\natexlab{b}}, A\&A, 434,
  483

\bibitem[{{Pietsch} {et~al.}(2007){Pietsch}, {Haberl}, {Sala}, {Stiele},
  {Hornoch}, {Riffeser}, {Fliri}, {Bender}, {B{\"u}hler}, {Burwitz}, {Greiner},
  \& {Seitz}}]{2007A&A...465..375P}
{Pietsch}, W., {Haberl}, F., {Sala}, G., {et~al.} 2007, A\&A, 465, 375

\bibitem[{{Primini} {et~al.}(1993){Primini}, {Forman}, \&
  {Jones}}]{1993ApJ...410..615P}
{Primini}, F.~A., {Forman}, W., \& {Jones}, C. 1993, ApJ, 410, 615

\bibitem[{{Stanek} \& {Garnavich}(1998)}]{1998ApJ...503L.131S}
{Stanek}, K.~Z. \& {Garnavich}, P.~M. 1998, ApJ, 503, L131

\bibitem[{{Stark} {et~al.}(1992){Stark}, {Gammie}, {Wilson}, {Bally}, {Linke},
  {Heiles}, \& {Hurwitz}}]{1992ApJS...79...77S}
{Stark}, A.~A., {Gammie}, C.~F., {Wilson}, R.~W., {et~al.} 1992, ApJS, 79, 77

\bibitem[{{Stiele} {et~al.}(2007{\natexlab{a}}){Stiele}, {Pietsch}, {Haberl},
  \& {Freyberg}}]{2007TV....M31....1S}
{Stiele}, H., {Pietsch}, W., {Haberl}, F., \& {Freyberg}, M.
  2007{\natexlab{a}}, submitted to A\&A

\bibitem[{{Stiele} {et~al.}(2007{\natexlab{b}}){Stiele}, {Pietsch}, {Haberl},
  {Freyberg}, \& {Trigo}}]{2007ATel.1191....1S}
{Stiele}, H., {Pietsch}, W., {Haberl}, F., {Freyberg}, M., \& {Trigo}, M.~D.
  2007{\natexlab{b}}, The Astronomer's Telegram, 1191, 1

\bibitem[{{Supper} {et~al.}(2001){Supper}, {Hasinger}, {Lewin}, {Magnier}, {van
  Paradijs}, {Pietsch}, {Read}, \& {Tr{\" u}mper}}]{2001A&A...373...63S}
{Supper}, R., {Hasinger}, G., {Lewin}, W.~H.~G., {et~al.} 2001, A\&A, 373, 63

\bibitem[{{Supper} {et~al.}(1997){Supper}, {Hasinger}, {Pietsch}, {Truemper},
  {Jain}, {Magnier}, {Lewin}, \& {van Paradijs}}]{1997A&A...317..328S}
{Supper}, R., {Hasinger}, G., {Pietsch}, W., {et~al.} 1997, A\&A, 317, 328

\bibitem[{{Trinchieri} \& {Fabbiano}(1991)}]{1991ApJ...382...82T}
{Trinchieri}, G. \& {Fabbiano}, G. 1991, ApJ, 382, 82

\bibitem[{{Trudolyubov} \& {Priedhorsky}(2007)}]{2007arXiv0708.0874T}
{Trudolyubov}, S. \& {Priedhorsky}, W. 2007, submitted to ApJ, astro-ph 0708.0874

\bibitem[{{Trudolyubov} {et~al.}(2006){Trudolyubov}, {Priedhorsky}, \&
  {Cordova}}]{2006ApJ...645..277T}
{Trudolyubov}, S., {Priedhorsky}, W., \& {Cordova}, F. 2006, ApJ, 645, 277

\bibitem[{{van Speybroeck} {et~al.}(1979){van Speybroeck}, {Epstein}, {Forman},
  {Giacconi}, {Jones}, {Liller}, \& {Smarr}}]{1979ApJ...234L..45V}
{van Speybroeck}, L., {Epstein}, A., {Forman}, W., {et~al.} 1979, ApJ, 234,
  L45

\bibitem[{{Williams}(2007)}]{2007Chan..M31....1W}
{Williams}, B.~F. 2007, in this MPE report

\bibitem[{{Williams} {et~al.}(2005){Williams}, {Garcia}, {McClintock},
  {Primini}, \& {Murray}}]{2005ApJ...632.1086W}
{Williams}, B.~F., {Garcia}, M.~R., {McClintock}, J.~E., {Primini}, F.~A., \&
  {Murray}, S.~S. 2005, ApJ, 632, 1086

\bibitem[{{Williams} {et~al.}(2006{\natexlab{a}}){Williams}, {Garcia},
  {McClintock}, {Primini}, \& {Murray}}]{2006ApJ...637..479W}
{Williams}, B.~F., {Garcia}, M.~R., {McClintock}, J.~E., {Primini}, F.~A., \&
  {Murray}, S.~S. 2006{\natexlab{a}}, ApJ, 637, 479

\bibitem[{{Williams} {et~al.}(2006{\natexlab{b}}){Williams}, {Naik}, {Garcia},
  \& {Callanan}}]{2006ApJ...643..356W}
{Williams}, B.~F., {Naik}, S., {Garcia}, M.~R., \& {Callanan}, P.~J.
  2006{\natexlab{b}}, ApJ, 643, 356

\end{thebibliography}
   

\end{document}